\documentclass[11pt]{JHEP3}
\usepackage{epsfig}
\usepackage{amssymb}
\usepackage{graphics}
\usepackage{bbm}
        \let\f=\phi

\let\F=\Phi           
\newcommand{\be}{\begin{equation}}
\newcommand{\ee}{\end{equation}}
\newcommand{\bea}{\begin{eqnarray}}
\newcommand{\eea}{\end{eqnarray}}
\newcommand{\ba}{\begin{array}}
\newcommand{\ea}{\end{array}}
\def\nn{\nonumber}

\title{Quark mass hierarchies in D-brane realizations of the Standard Model}
\author{P. Anastasopoulos$^{1,2}$\footnote{Pascal.Anastasopoulos@roma2.infn.it} 
and Andrea Lionetto$^{1}$\footnote{Andrea.Lionetto@roma2.infn.it}\\
~\\
$^1$ Dip. di Fisica \& Sez. I.N.F.N. \ Universit{\`a} di
Roma \ ``Tor Vergata'',\\
Via della Ricerca Scientifica, 1 - 00133 \ Roma, \ ITALY\\
~\\
$^2$ Department of Physics, CERN-Theory Division,\\ 
1211 Geneva 23, SWITZERLAND}

\preprint{0912.0121[hep-ph] \\ 
ROM2F/2009/26\\
CERN-PH-TH/2009-235
}

\abstract{This proceeding is based on arXiv:0905.3044. 
We analyze the problem of the hierarchy of masses and mixings in orientifold 
realizations of the Standard Model. We present a bottom-up brane configuration that can generate such hierarchies.}

\begin{document}

\section{Introduction, motivation  and scope}

One of the biggest puzzles in the Standard Model (SM) is the origin and hierarchy
of masses and mixings. When it comes to masses the scale of the problem is enormous: one needs to
explain a range of masses that spans fifteen orders of magnitude between the mass of the lightest
neutrino to the top quark. Moreover the pattern of mixings is interesting. In the quark sector the first and second family mix strongly while all other mixings are small. In the lepton sector, all mixings measured so far are maximal. It seems to suggest that as we move up in mass mixings tend to become smaller.

There are several ideas on the origin of mass, which can be roughly lumped into four classes: radiative mechanisms, texture zeros, family symmetries
and seesaw mechanisms. However, the classes are not
completely disjoint. In particular texture zeros can be considered as a class of family symmetries as
they are usually implemented via a discrete symmetry. 


In this work we analyze SM mass patterns and hierarchies in the context
of open string theory vacua, alias orientifolds~\cite{Bianchi:1989du}.
These vacua allow for a small string scale $M_s$ and provided a fresh new perspective in the search for the SM \cite{Uranga:2003pz}. Moreover they allowed a bottom-up approach \cite{Antoniadis:2000ena, Aldazabal:2000sa} to building the SM, by utilizing the geometrized language offered by D-branes supporting the SM interactions and particles.

In this framework, the SM particles are strings which are stretched between some
stacks of D-branes (aka the ``SM branes"). Typically there is a need of some extra stacks 
(aka the ``hidden sector") which do not include light observable-hidden strings \cite{Dijkstra:2004cc}.


Anomalous U(1) symmetries are ubiquitous in orientifolds \cite{Anastasopoulos:2006cz}. 
It has been
argued early on \cite{Uranga:2003pz, Antoniadis:2000ena}, that any 
SM orientifold realization must have at least one
and generically three anomalous U(1) symmetries, that make the most
characteristic signature of orientifold vacua. Their phenomenological implications 
are diverse~\cite{Kiritsis:2002aj}.

Their most important property, that impacts importantly on the dynamics of
the D-brane stack is that they provide numerous selection rules on the
effective couplings. In particular, they may be responsible for the absence
of the $\mu$-term, Yukawa couplings, baryon and lepton violating couplings etc.
However, as anomalous U(1)'s are effectively broken as gauge symmetries, the
selection rules they provide need qualification.
As the breaking of the gauge symmetry happens via the mixing with RR forms,
the global U(1) symmetry remains at this stage intact.
There are two types of realizations of anomalous U(1) symmetries as global
symmetries. If D-terms force charged fields to obtain vev's
then the global U(1) symmetry is broken. If on the other hand no vev's are
generated the anomalous U(1) global symmetry remain intact in perturbation theory.


However, the story must change beyond perturbation theory for two reasons.
The first is that we do not expect exact (compact) global symmetries to survive
in a gravitational theory. The second (in agreement with the first) is that there are always non-perturbative
effects that violate the associated global symmetry. The argument is simple.
A U(1) transformation involves a shift of RR field. The associated D-instanton effect
which is charged under the same RR field (the Stuckelberg axion) will violate by definition
the associated global U(1) symmetry. 
The effect is a D-instanton effect, whose
field theory limit sometimes may admit a gauge instanton interpretation~\cite{Bachas:1997mc}.
\FIGURE[t]{\epsfig{file=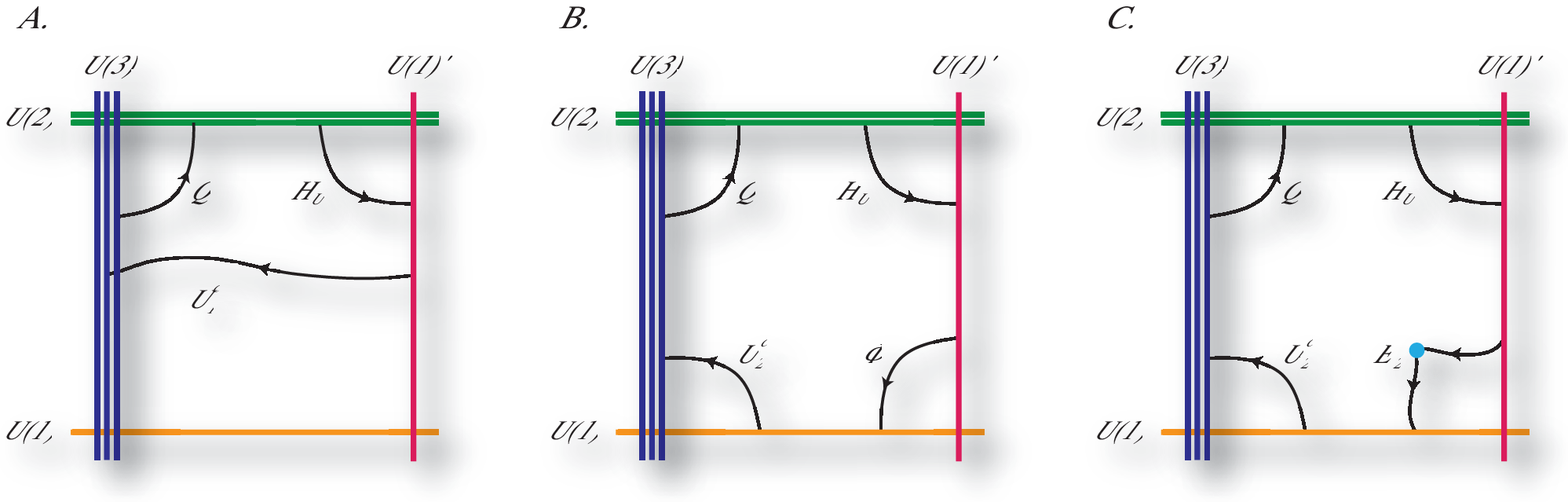,width=15cm}\caption{The three types of
mass generating terms: The configuration A allows for a Yukawa term. However, in the B and C cases
no Yukawa terms can be generated. In the B case there is a higher order term due to
the presence of a field $\F$, while in the C case there is a contribution from an instanton term $E_2$. \label{Picture}}}

Therefore, there are several effects that can produce  hierarchically different Yukawa-like couplings
(figure \ref{Picture})\footnote{We would like to mention that higher order or instantonic terms might bring in the potential unwanted couplings too. By ``unwanted couplings" we mention all terms that could make the model phenomenologically uninteresting, like baryon or lepton number violating terms or couplings that lead to fast proton decay. One can easily assume that such couplings are not present due to the absence of the related instanton. However, if an unwanted coupling in the superpotential has the same violation of U(1) charges as a mass term that has a non-zero instanton contribution, then there is a non-trivial instanton contribution for this term and thus it will generated with a similar strength. Such models must be excluded.}
\cite{Leontaris:2009ci}-\cite{Cvetic:2009yh}.
\begin{itemize}
\item Tree-level cubic Yukawa couplings: This is the generic case when such
couplings are allowed. Their coefficient depends in general on several
ingredients. It is always proportional to the ten-dimensional dilaton but
also internal volumes, and other
backgrounds fields (internal magnetic fields, fluxes) enter. They may be
correlated with the associated gauge couplings if the fields participating
come from overlapping D-branes. They may also be free of volumes if the branes
intersect at points. Such variations are enough some times to explain the 
mass hierarchy inside a family \cite{Antoniadis:2000ena}.
\item Higher order couplings: These are couplings that appear beyond the cubic
level. They necessarily involve more fields than the SM fields.
These extra fields must obtain an expectation value in order for an effective
Yukawa coupling  to be generated.
Then such couplings  compared to the previous case carry an extra factor of
$\left({\langle\phi\rangle/ M_s}\right)^n$ with $n$ a positive integer.
Depending on the compactification the string scale may be replaced by a
compactification scale.
If  $\langle\phi\rangle\ll M_s$ this generates a hierarchy in the associated
Yukawa coupling. 
\item D-Instanton-generated couplings: Such couplings violate the anomalous U(1)
symmetries. They are suppressed by exponential instanton factors
of the form $e^{-1/g}$ where $g$ is linearly related to the ten-dimensional
coupling constant and depends also on the volume of the cycle
the D-instanton is wrapped-on, as well as on magnetic fields, fluxes, etc.
In the particular case of gauge instantons $g$ is the square of the associated
gauge coupling.
In the well-controlled regime, $g\ll 1$ and multi-instantons are suppressed.
Beyond the instanton-action factor, instanton-generated couplings carry a
characteristic scale. This is determined by the string scale, or other volume factors
affecting the world-volume factor of the D-instanton. Finally there is a one-loop
determinant that is generically of order ${\cal O}$(1).
\item Not generated couplings: Remain zero as no vev or instanton can generate them.
\end{itemize}
In \cite{Anastasopoulos:2009mr}, we explore different effects that are prone to generate
interesting hierarchies between fermion masses. 
In particular, for four brane stacks we found vacua in which for each fermionic mass matrix, the highest mass scale is related to Yukawa terms, the intermediate mass scale to instantons while the lowest scale to higher order terms. The CKM matrix computed for this model agrees with the experimental result. 
We would like to mention that our study was based on the bottom-up approach.
The goal was to identify D-brane configurations that are promising when it comes
to generating the fermion hierarchy. The next step will be to construct such interesting
D-bane configurations.

\section{An example of the new hierarchy idea}


In order to show our results we focus on a vacuum with four stacks of branes: a stack of three (color stack), a stack of two (weak stack), and two single branes. The hypercharge embedding is $Y={1\over 6} Q_3+{1\over 2} Q_1+{1\over 2} Q_1'$. The SM particles are:
\bea
\begin{array}{lllllllllll}
Q_1~~~~&:& ~(~~V, ~~V, ~~0, ~~0) ~~~&,&~~ Q_2,Q_3~&:&~(~~V, ~~ \bar V ,~~ 0 ,~~ 0)\\
U^c_1~~~~&:&~(~~ \bar V, ~~0 , ~~ \bar V, ~~0) ~~~&,&~~ U^c_2~U^c_3~&:&~(~~ \bar V,~~ 0 , ~~0 , ~~ \bar V)\\
D^c_1~~~~&:&~(~~ \bar V, ~~0 , ~~V, ~~0)~~~&,&~~ D^c_2~D^c_3~&:&~(~~ \bar V, ~~0 ,~~ 0 , ~~V)\\
L^c_1~~~~&:&~(~~0, ~~V , ~~ \bar V , ~~0) ~&,&~~ L^c_2~L^c_3~&:&~(~~0, ~~V, ~~0,~~ \bar V)\\
E^c_1~~~~&:&~(~~0, ~~0 , ~~0 , ~~S) ~~~&,&~~ E^c_2~~~~~&:&~(~~0, ~~0 , ~~0, ~~S) ~~&,&~~ E^c_2~&:&~(~~0, ~~0 ,~~ S, ~~0)\\
N^c_{1,2,3}~&:&~(~~0,~~ A , ~~0, ~~0)\\
H_u~&:&~ (~~0, ~~ \bar V , ~~V , ~~0) ~~~&,&~~ H_d~&:&~ (~~0, ~~ \bar V ,~~ \bar V , ~~0)
\end{array} \label{modelo1}\eea
where $V,~A,~S$ are the fundamental, antisymmetric and symmetric representation of the corresponding 
group with U(1) charges: $V\to 1,~A\to 2,~S\to 2$. All bars denote the conjugates rep with opposite U(1) charge.
We also consider two additional scalars with zero hypercharge $\f_1$ and $\f_2$, coming from the non chiral part of the spectrum:
\bea
&&\f_1~:~ (0, 0 , ~~ \bar V , V)  ~~~,~~ \f_2~:~ (0, 0 , V , ~~ \bar V) ~.
\eea
The above spectrum satisfies all irreducible anomaly cancellation conditions and additionally some extra conditions that are coming from tadpole cancellation.

For the above spectrum (\ref{modelo1}): $Q_1$, $U^c_1$, and $H_u$ form a perturbative Yukawa term $Q_1 U^c_1 H_u \to m_{1,1}\sim \langle  H_u \rangle$. On the other hand $Q_1$, $U^c_{2,3}$, and $H_u$ do not allow for such term. 
The contribution of the extra field $\f_1$ provides a higher term $Q_1 U^c_1 H_u \f_1 \to m_{1,2}\sim \langle  H_u \rangle \langle  \f_1 \rangle /M_s$.
For the rest of the entries in the mass matrix, neither $\f_{1,2}$ can form gauge invariant couplings, thus we assume the presence of instantons, like: $Q_2 U^c_1 H_u \to m_{2,1} \sim \langle  H_u \rangle e^{-Vol}$.
Following this spirt we evaluate the corresponding mass matrices for the quarks:
\bea
&&M_U =\langle{H_u}\rangle~\left(
\begin{array}{lll}
g_1 ~        &g_2 v_{\f_1} &g_3 v_{\f_1} \\
g_4 E_1 &g_5 E_2        &g_6 E_2 \\
g_7 E_1 &g_8 E_2        &g_9 E_2 \\
\end{array}\right)
~~~,~~ 
M_D =\langle{H_d}\rangle~\left(
\begin{array}{lll}
q_1 ~        &q_2 v_{\f_2} &q_3 v_{\f_2} \\
q_4 E_1 &q_5 E_3       &q_6 E_3 \\
q_7 E_1 &q_8 E_3       &q_9E_3 \\
\end{array}\right)\label{Massmatrices}\\
%
&&M_L =\langle{H_d}\rangle~\left(
\begin{array}{lll}
l_1 E_4 &l_2 v_{\f_1}  &l_3 ~ \\
l_4 E_4 &l_5 v_{\f_1}  &l_6 ~ \\
l_7 E_4 &l_8 v_{\f_1}  &l_9 ~ \\
\end{array}\right)\nn\\
&&M_N =
\left(
\begin{array}{llllll}
0 & 0 & 0 & g_{11} \langle{H_u}\rangle E_1 & g_{12} \langle{H_u}\rangle E_1 & g_{13} \langle{H_u}\rangle E_1 \\
0 & 0 & 0 & g_{21} \langle{H_u}\rangle E_1 & g_{22} \langle{H_u}\rangle E_1 & g_{23} \langle{H_u}\rangle E_1 \\
0 & 0 & 0 & g_{31} \langle{H_u}\rangle E_1 & g_{32} \langle{H_u}\rangle E_1 & g_{33} \langle{H_u}\rangle E_1 \\
g_{11} \langle{H_u}\rangle E_1 & g_{21} \langle{H_u}\rangle E_1 & g_{31} \langle{H_u}\rangle E_1 & q_{11}M_sE_5 & q_{12}M_sE_5 & q_{13}M_sE_5 \\
g_{12} \langle{H_u}\rangle E_1 & g_{22} \langle{H_u}\rangle E_1 & g_{32} \langle{H_u}\rangle E_1  & q_{21}M_sE_5 & q_{22}M_sE_5 & q_{23}M_sE_5 \\
g_{13} \langle{H_u}\rangle E_1 & g_{23} \langle{H_u}\rangle E_1 & g_{33} \langle{H_u}\rangle E_1  & q_{31}M_sE_5 & q_{32}M_sE_5 & q_{33}M_sE_5
\end{array}\right)\nn
\eea
where $g_i,~q_i,~l_i$, $g_{ij}$ and $q_{ij}$ are dimensionless couplings assumed
to be of the same order $[0.1-0.6]$.
Also $v_{\f_1}=\langle{\f_1}\rangle/M_s$, ${v_\f}_2=\langle{\f_2}\rangle/M_s$ and $E_i=e^{-Vol_{I_i} I_i}$ are the dimensionless instantons.

For the present vacuum we were able to find solutions where
there is 1-1 correspondence between the fermion masses in each 
family and the Yukawa, higher order and instantonic terms:
\bea
\left.
\begin{array}{lllll}
\langle{H_u}\rangle\sim m_t&,&~~~~~~~~\langle{H_d}\rangle\sim m_b\\
E_1\sim E_2 \sim m_c/m_t
&,&~~~~~~~~ E_3\sim E_4\sim m_s/m_b ~&,&E_{5}\sim 0.654\\
v_{\phi_1} \sim m_u/m_t
&, &~~~~~~~~v_{\phi_2} \sim m_d/m_b
\end{array}\right.
\label{YIH}\eea
where $m_i$ are the masses of the corresponding quarks \cite{Xing:2007fb},
and all couplings $|g_i|$, $|q_i|$, $|l_i|$, $|g_{ij}|$, $|q_{ij}|$ are within the range $[0.1,0.6]$.
Notice that $E_5$ appears always multiplied by $M_s$ and it's value triggers the seesaw mechanism. It is
the only value that changes if we repeat the same mechanism at other scales apart 
from 1 TeV.

Using the above values for the couplings and vev's, we can proceed and evaluate the Cabibbo - Kobayashi - Maskawa Matrix (CKM). For the above vacuum, the matrix is:
\bea
{\rm CKM(1\, TeV)} &=&
\left(
\begin{array}{lll}
0.970 & 0.240 & 0.007 \\
0.240 & 0.970 & 0.013 \\
0.010 & 0.011 & 0.999
\end{array}
\right)\eea
that has to be compared with the experimental data \cite{Amsler:2008zzb}:
\bea
{\rm CKM(Data)} &=&
\left(
\begin{array}{rrr}
0.97419 \pm 0.00022 & 0.2257\pm 0.0010      & ~~~~~0.00359 \pm 0.00016\\
0.2256 \pm 0.0010      & ~~~~~0.97334\pm 0.00023 & 0.0415 \pm 0.001 \\
0.00874^{+0.00026}_{- 0.00037}
                                        & 0.0407\pm 0.0010      &
0.999133^{+0.000044}_{- 0.000043}
\end{array}
\right)
~~~~~~~ \label{CKMdata}\eea

In this vacuum, only the $\mu$-term and same specific Yukawa couplings share the same instanton $E_1$ and therefore $\mu \sim E_1 \sim m_c/m_t$. This is not problematic for low string scale models.

\section{Conclusions}

In the field theory context several ideas has been proposed to explain the fermionic mass hierarchy in the SM such as radiative mechanisms, texture zeros, family symmetries and seesaw mechanisms. None of them can be separately implemented in D-brane constructions whose aim is to reproduce the SM. New types of textures arise due to extended symmetries, i.e. several (anomalous) $U(1)$'s.  

In this proceeding we argued that SM mass hierarchies can come from different kind of terms that are generically present in D-brane vacua. Such terms are the perturbative Yukawas, higher order terms (with the contribution of an extra scalar field) and instantonic terms. Hierarchies are due to the 1-1 correspondence between the masses of the fermions in each mass matrix with the three terms mentioned above.

In this framework we have shown a model that materialize this idea. However the model is constructed on the bottom-up fashion and a consistent open string vacuum has to be still constructed.

\centerline{\bf Acknowledgements}
\addcontentsline{toc}{section}{Acknowledgements}

We would especially like to thank Elias Kiritsis for the fruitful collaboration on the paper 
where this proceeding is based on.
We would like also to thank the organizers of Corfu 2009 for giving us 
the opportunity to present this work.


\begin{thebibliography}{10}






\bibitem{Bianchi:1989du}
 M.~Bianchi and A.~Sagnotti,
 Phys.\ Lett.\  B {\bf 231}, 389 (1989).
%
%
%
%
 G.~Pradisi and A.~Sagnotti,
 Phys.\ Lett.\  B {\bf 216}, 59 (1989).
%
%
 C.~Angelantonj and A.~Sagnotti,
 Phys.\ Rept.\  {\bf 371}, 1 (2002)
 [arXiv:hep-th/0204089].

\bibitem{Uranga:2003pz}
 A.~M.~Uranga,
 Class.\ Quant.\ Grav.\  {\bf 20}, S373 (2003)
 [arXiv:hep-th/0301032].
%
 E.~Kiritsis,
 Fortsch.\ Phys.\  {\bf 52}, 200 (2004)
 [arXiv:hep-th/0310001].
%
 R.~Blumenhagen, M.~Cvetic, P.~Langacker and G.~Shiu,
 Ann.\ Rev.\ Nucl.\ Part.\ Sci.\  {\bf 55}, 71 (2005)
 [arXiv:hep-th/0502005].
%

\bibitem{Antoniadis:2000ena}
%
%
 I.~Antoniadis, E.~Kiritsis, J.~Rizos and T.~N.~Tomaras,
 Nucl.\ Phys.\  B {\bf 660}, 81 (2003)
 [arXiv:hep-th/0210263].

\bibitem{Aldazabal:2000sa}
 G.~Aldazabal, L.~E.~Ibanez, F.~Quevedo and A.~M.~Uranga,
 JHEP {\bf 0008}, 002 (2000)
 [arXiv:hep-th/0005067].

\bibitem{Dijkstra:2004cc}
 T.~P.~T.~Dijkstra, L.~R.~Huiszoon and A.~N.~Schellekens,
 Nucl.\ Phys.\  B {\bf 710}, 3 (2005)
 [arXiv:hep-th/0411129].
%
 P.~Anastasopoulos, T.~P.~T.~Dijkstra, E.~Kiritsis and A.~N.~Schellekens,
 Nucl.\ Phys.\  B {\bf 759}, 83 (2006)
 [arXiv:hep-th/0605226].






\bibitem{Anastasopoulos:2006cz}
%
 I.~Antoniadis, E.~Kiritsis and J.~Rizos,
 Nucl.\ Phys.\  B {\bf 637}, 92 (2002)
 [arXiv:hep-th/0204153].
%
 P.~Anastasopoulos, M.~Bianchi, E.~Dudas and E.~Kiritsis,
 JHEP {\bf 0611}, 057 (2006)
 [arXiv:hep-th/0605225].
%
%
 P.~Anastasopoulos,
 JHEP {\bf 0308}, 005 (2003)
 [arXiv:hep-th/0306042].
%
 Phys.\ Lett.\  B {\bf 588}, 119 (2004)
 [arXiv:hep-th/0402105].
%
 Int.\ J.\ Mod.\ Phys.\  A {\bf 22}, 5808 (2008).
%
 J.\ Phys.\ Conf.\ Ser.\  {\bf 53}, 731 (2006).
%
 Fortsch.\ Phys.\  {\bf 55}, 633 (2007)
 [arXiv:hep-th/0701114].

\bibitem{Kiritsis:2002aj}
 E.~Kiritsis and P.~Anastasopoulos,
 JHEP {\bf 0205}, 054 (2002)
 [arXiv:hep-ph/0201295].
%
 D.~M.~Ghilencea, L.~E.~Ibanez, N.~Irges and F.~Quevedo,
 JHEP {\bf 0208}, 016 (2002)
 [arXiv:hep-ph/0205083].
%
 B.~Kors and P.~Nath,
 Phys.\ Lett.\  B {\bf 586}, 366 (2004)
 [arXiv:hep-ph/0402047].
%
 C.~Coriano', N.~Irges and E.~Kiritsis,
 Nucl.\ Phys.\  B {\bf 746}, 77 (2006)
 [arXiv:hep-ph/0510332].
%
%
 P.~Anastasopoulos, F.~Fucito, A.~Lionetto, G.~Pradisi, A.~Racioppi and Y.~S.~Stanev,
 Phys.\ Rev.\  D {\bf 78}, 085014 (2008)
 [arXiv:0804.1156 [hep-th]].
%
 E.~Dudas, Y.~Mambrini, S.~Pokorski and A.~Romagnoni,
 JHEP {\bf 0908}, 014 (2009)
 [arXiv:0904.1745 [hep-ph]].
%
 I.~Antoniadis, A.~Boyarsky, S.~Espahbodi, O.~Ruchayskiy and J.~D.~Wells,
 Nucl.\ Phys.\  B {\bf 824}, 296 (2010)
 [arXiv:0901.0639 [hep-ph]].

\bibitem{Bachas:1997mc}
 C.~Bachas, C.~Fabre, E.~Kiritsis, N.~A.~Obers and P.~Vanhove,
 Nucl.\ Phys.\  B {\bf 509}, 33 (1998)
 [arXiv:hep-th/9707126].
%
 E.~Kiritsis,
 arXiv:hep-th/9906018.
%
 M.~Billo, M.~Frau, I.~Pesando, F.~Fucito, A.~Lerda and A.~Liccardo,
 JHEP {\bf 0302}, 045 (2003)
 [arXiv:hep-th/0211250].
%
 R.~Blumenhagen, M.~Cvetic and T.~Weigand,
 Nucl.\ Phys.\  B {\bf 771}, 113 (2007)
 [arXiv:hep-th/0609191].
%
 M.~Haack, D.~Krefl, D.~Lust, A.~Van Proeyen and M.~Zagermann,
 JHEP {\bf 0701}, 078 (2007)
 [arXiv:hep-th/0609211].
%
 L.~E.~Ibanez and A.~M.~Uranga,
 JHEP {\bf 0703}, 052 (2007)
 [arXiv:hep-th/0609213].
%
 B.~Florea, S.~Kachru, J.~McGreevy and N.~Saulina,
 JHEP {\bf 0705}, 024 (2007)
 [arXiv:hep-th/0610003].
%
%
 M.~Bianchi and E.~Kiritsis,
 Nucl.\ Phys.\  B {\bf 782}, 26 (2007)
 [arXiv:hep-th/0702015].
%
%
 M.~Bianchi, F.~Fucito and J.~F.~Morales,
 JHEP {\bf 0707}, 038 (2007)
 [arXiv:0704.0784 [hep-th]].
%
 L.~E.~Ibanez, A.~N.~Schellekens and A.~M.~Uranga,
 JHEP {\bf 0706}, 011 (2007)
 [arXiv:0704.1079 [hep-th]].
%
 M.~Cvetic, R.~2.~Richter and T.~Weigand,
 Phys.\ Rev.\  D {\bf 76}, 086002 (2007)
 [arXiv:hep-th/0703028].
%
 M.~Cvetic and T.~Weigand,
 Phys.\ Rev.\ Lett.\  {\bf 100}, 251601 (2008)
 [arXiv:0711.0209 [hep-th]].
%
 R.~Blumenhagen, M.~Cvetic, D.~Lust, R.~2.~Richter and T.~Weigand,
 Phys.\ Rev.\ Lett.\  {\bf 100}, 061602 (2008)
 [arXiv:0707.1871 [hep-th]].
%
 L.~E.~Ibanez and R.~2.~Richter,
 JHEP {\bf 0903}, 090 (2009)
 [arXiv:0811.1583 [hep-th]].
%
 C.~Angelantonj, C.~Condeescu, E.~Dudas and M.~Lennek,
 Nucl.\ Phys.\  B {\bf 818}, 52 (2009)
 [arXiv:0902.1694 [hep-th]].

\bibitem{Cremades:2003qj}
 D.~Cremades, L.~E.~Ibanez and F.~Marchesano,
 JHEP {\bf 0307}, 038 (2003)
 [arXiv:hep-th/0302105].

\bibitem{Ibanez:2001nd}
 L.~E.~Ibanez, F.~Marchesano and R.~Rabadan,
 JHEP {\bf 0111}, 002 (2001)
 [arXiv:hep-th/0105155].


\bibitem{Xing:2007fb}
 Z.~z.~Xing, H.~Zhang and S.~Zhou,
 Phys.\ Rev.\  D {\bf 77}, 113016 (2008)
 [arXiv:0712.1419 [hep-ph]].

\bibitem{Leontaris:2009ci}
 G.~K.~Leontaris,
 arXiv:0903.3691 [hep-ph].
 G.~K.~Leontaris and N.~D.~Vlachos,
 arXiv:0909.4701 [hep-th].

%
\bibitem{Kiritsis:2008ry}
 E.~Kiritsis, B.~Schellekens and M.~Tsulaia,
 JHEP {\bf 0810}, 106 (2008)
 [arXiv:0809.0083 [hep-th]].
 E.~Kiritsis, M.~Lennek and B.~Schellekens,
 arXiv:0909.0271 [hep-th].

\bibitem{Anastasopoulos:2009mr}
 P.~Anastasopoulos, E.~Kiritsis and A.~Lionetto,
 JHEP {\bf 0908} (2009) 026
 [arXiv:0905.3044 [hep-th]].



\bibitem{Cvetic:2009yh}
{ M.~Cvetic, J.~Halverson and R.~Richter},
 [arXiv:{0905.3379}{[hep-th]}].
%
 M.~Cvetic, J.~Halverson and R.~Richter,
 arXiv:0909.4292 [hep-th].
%
 M.~Cvetic, J.~Halverson and R.~Richter,
 arXiv:0910.2239 [hep-th].
%
 M.~Cvetic, I.~Garcia-Etxebarria and R.~Richter,
 arXiv:0911.0012 [hep-th].




\bibitem{Amsler:2008zzb}
 C.~Amsler {\it et al.}  [Particle Data Group],
 Phys.\ Lett.\  B {\bf 667}, 1 (2008).







\end{thebibliography}
\end{document}